\documentclass[prb,superscriptaddress,reprint]{revtex4-2}

\usepackage{graphicx}
\usepackage{dcolumn}
\usepackage{bm}
\usepackage{gensymb}
\usepackage{mathptmx} 
\usepackage{siunitx} 
\usepackage{xcolor}

\begin{document}




\title{Role of element-specific damping on the ultrafast, helicity-independent all-optical switching dynamics in amorphous (Gd,Tb)Co thin films.}


\author{Alejandro Ceballos}
\thanks{These two authors contributed equally}
\affiliation{Department of Materials Science and Engineering, University of California Berkeley, Berkeley,
California 94720, USA}
\affiliation{Materials Sciences Division, Lawrence Berkeley National Laboratory, Berkeley, California 94720, USA}
    
\author{Akshay Pattabi}
\thanks{These two authors contributed equally}
	\affiliation{Department of Electrical Engineering and Computer Sciences, University of California, Berkeley, Berkeley, CA 94720, USA}
    
\author{Amal El-Ghazaly}
	\affiliation{Department of Electrical Engineering and Computer Sciences, University of California, Berkeley, Berkeley, CA 94720, USA}
\author{Sergiu Ruta}
	\affiliation{Department of Physics, University of York, Heslington, York YO10 5DD, United Kingdom}
\author{Christian P Simon}
    \affiliation{Department of Physics, University of California, Berkeley, Berkeley, California 94720, USA}
\author{Richard F L Evans}
    \affiliation{Department of Physics, University of York, Heslington, York YO10 5DD, United Kingdom}
\author{Thomas Ostler}
    \affiliation{Materials and Engineering Research Institute, Sheffield Hallam University, Howard Street, Sheffield S1 1WB, UK}
    \affiliation{Faculty of Science, Technology and Arts, Sheffield Hallam University, Howard Street, Sheffield, S1 1WB, UK}
\author{Roy W Chantrell}
    \affiliation{Department of Physics, University of York, Heslington, York YO10 5DD, United Kingdom}
\author{Ellis Kennedy}
    \affiliation{Department of Materials Science and Engineering, University of California Berkeley, Berkeley,
California 94720, USA}
    \affiliation{National  Center  for  Electron  Microscopy, Molecular  Foundry, Lawrence  Berkeley  National  Laboratory,  Berkeley, CA 94720}
\author{Mary Scott}
    \affiliation{Department of Materials Science and Engineering, University of California Berkeley, Berkeley,
California 94720, USA}
    \affiliation{National  Center  for  Electron  Microscopy, Molecular  Foundry, Lawrence  Berkeley  National  Laboratory,  Berkeley, CA 94720}
\author{Jeffrey Bokor}
    \affiliation{Department of Electrical Engineering and Computer Sciences, University of California, Berkeley, Berkeley, CA 94720, USA}
	\affiliation{Materials Sciences Division, Lawrence Berkeley National Laboratory, Berkeley, California 94720, USA}
\author{Frances Hellman}
    \affiliation{Department of Materials Science and Engineering, University of California Berkeley, Berkeley,
    California 94720, USA}
	\affiliation{Materials Sciences Division, Lawrence Berkeley National Laboratory, Berkeley, California 94720, USA}
    \affiliation{Department of Physics, University of California, Berkeley, Berkeley, California 94720, USA}


\begin{abstract}

Ultrafast control of the magnetization in ps timescales by fs laser pulses offers an attractive avenue for applications such as fast magnetic devices for logic and memory. However, ultrafast helicity-independent all-optical switching (HI-AOS) of the magnetization has thus far only been observed in Gd-based, ferrimagnetic amorphous (\textit{a}-) rare earth-transition metal (\textit{a}-RE-TM) systems, and a comprehensive understanding of the reversal mechanism remains elusive. Here, we report HI-AOS in ferrimagnetic \textit{a}-Gd$_{22-x}$Tb$_x$Co$_{78}$ thin films, from x = 0 to x = 18, and elucidate the role of Gd in HI-AOS in \textit{a}-RE-TM alloys and multilayers. Increasing Tb content results in increasing perpendicular magnetic anisotropy and coercivity, without modifying magnetization density, and slower remagnetization rates and higher critical fluences for switching but still shows picosecond HI-AOS. Simulations of the atomistic spin dynamics based on the two-temperature model reproduce these results qualitatively and predict that the lower damping on the RE sublattice arising from the small spin-orbit coupling of Gd (with $L = 0$) is instrumental for the faster dynamics and lower critical fluences of the Gd-rich alloys. Annealing \textit{a}-Gd$_{10}$Tb$_{12}$Co$_{78}$ leads to slower dynamics which we argue is due to an increase in damping. These simulations strongly indicate that acounting for element-specific damping is crucial in understanding HI-AOS phenomena. The results suggest that engineering the element specific damping of materials can open up new classes of materials that exhibit low-energy, ultrafast HI-AOS.
\end{abstract}

\maketitle

\section{Introduction}

The ability to control magnetism at short ps and sub-ps timescales has tantalized scientists since the discovery in 1996 of the ultrafast demagnetization of Ni  following irradiation by fs laser pulses,\cite{beaurepaire_ultrafast_1996} opening up the field of ultrafast magnetization dynamics. A major breakthrough was the discovery of helicity-independent all-optical switching (HI-AOS) of the magnetization, sometimes referred to as thermally-induced magnetization switching (TIMS), in ferrimagnetic \textit{a}-Gd-Fe-Co alloys\cite{stanciu_subpicosecond_2007, radu_transient_2011, ostler_ultrafast_2012} in ps timescales by a single fs laser pulse. This magnetization switching phenomenon is unique in the field of all-optical switching as it exhibits one of the fastest switching speeds recorded\cite{Yang_SciAdv_2018}, it is reversible and cyclable for 1000’s of repetitions, and it is thermally-driven by a single laser pulse irrespective of polarization, i.e. no transfer of angular momentum from the laser is involved. HI-AOS offers the possibility of promising technological applications in high-speed, energy-efficient and non-volatile magnetic memory and logic, with two to three orders of magnitude higher operating speeds compared to conventional spintronic devices that operate on mechanisms such as external field control,\cite{Gerrits_Nature_2002} spin-transfer- torque,\cite{Rowlands_APL_2011, Zhao_JPhysD_2011} or spin-orbit torque.\cite{Garello_APL_2014}

Despite significant advances in the field, a complete understanding of the mechanism of HI-AOS still remains lacking. Thus far, deterministic ultrafast toggle switching of the magnetization by a single laser pulse -- where the switched area is controlled just by local heating from the spatial distribution of the laser pulse fluence -- is predominantly found in Gd-based \textit{a}-RE-TM ferrimagnetic alloys and multilayers. These include \textit{a}-Gd-Fe-Co,\cite{stanciu_subpicosecond_2007,radu_transient_2011,ostler_ultrafast_2012}  \textit{a}-Gd-Co,\cite{elGhazaly_2019} Pt/Co/Gd,\cite{lalieu_deterministic_2017} and exchange coupled Co/Pt/Co/\textit{a}-GdFeCo\cite{gorchon_single_2017} systems. Similar Tb-based ferrimagnetic systems such as \textit{a}-Tb-Co alloys\cite{mangin_engineered_2014} have so far only shown helicity-dependent AOS (HD-AOS), requiring multiple circularly polarized laser pulses over longer timescales ($\mu$s to ms), \cite{MSElHadri_2016} or transient reversal with a single laser pulse, \cite{alebrand_subpicosecond_2014} wherein the magnetization reverts back to its original direction after a short reversal of a few ps. However, single-shot HI-AOS was demonstrated in Tb/Co multilayers when the layer thickness ratio of Co and Tb was within 1.3 - 1.5,\cite{aviles2019integration} although the switching dynamics were not reported. HI-AOS was also demonstrated in an \textit{a}-Tb$_{22}$Fe$_{69}$Co$_{9}$ alloy,\cite{liu_nanoscale_2015} but required patterning of nanoscale antennas to enhance the optical field, thereby confining the switched region to less than 100 nm in areas near and around the antennas. The switching in that experiment was strongly influenced by inhomogeneities and control of the uniformity in the switched area under the laser pulse fluence profile could not be achieved.

\begin{figure*}[ht!]
\centering
\includegraphics[width = 0.8\textwidth]{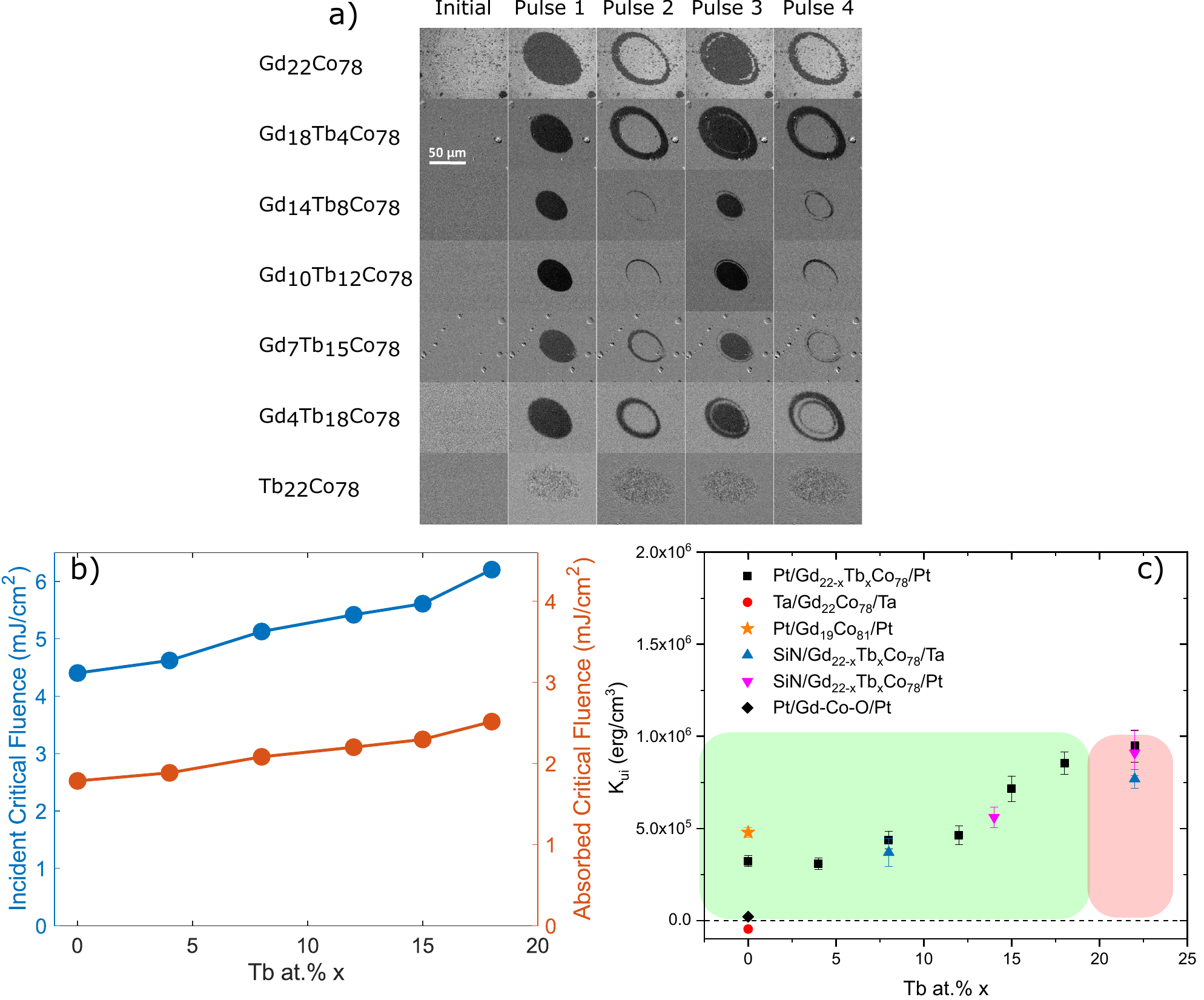}
  \caption{a) MOKE microscopy images of Ta(3nm)/Pt(3)/\textit{a}-Gd$_{22-x}$Tb$_x$Co$_{78}$(10)/Pt(3) thin films following a series of single laser pulses at an incident fluence of 6.9 mJ/cm$^2$, illustrate samples' ability to all-optically reverse their magnetization upon irradiation.
  Samples with a Tb concentration of up to 18 at.\% exhibited HI-AOS while \textit{a}-Tb$_{22}$Co$_{78}$ only exhibited demagnetization as evidenced by the nucleation of random domains. b) Incident and absorbed critical fluence increase with  increasing Tb content. c) Intrinsic anisotropy constant K$_{ui}$ vs Tb atomic percent in \textit{a}-Gd$_{22-x}$Tb$_x$Co$_{78}$ thin films with various capping and underlayers as shown in the legend. Anisotropy increases with increasing Tb content for fixed under and overlayers. Pt under or overlayers increased K$_{ui}$ for fixed x. The green box shows the compositions that exhibited HI-AOS; the red box those that did not. The effect of growing on different substrates and capping layers was tested but it had no effect on HI-AOS. Four \textit{a}-Gd-Co samples with varying Gd/Co ratio are shown at x = 0. At the top, indicated with an orange star, is \textit{a}-Gd$_{19}$Co$_{81}$ with compensation T below RT exhibiting PMA and HI-AOS. Beneath it the black square \textit{a}-Gd$_{22}$Co$_{78}$ which is part of the main study; it has Pt under and over layers and exhibits PMA and HI-AOS. The black diamond is Gd-Co-O also with Pt over and underlayers, but grown in a reactive oxygen atmosphere, with K$_{ui}$ value of 2.05 x 10$^4$ erg/cm$^{3}$ and M$_S$ of 40 emu/cm$^{3}$. It exhibited HI-AOS though its dynamics were not studied. In red is \textit{a}-Gd$_{22}$Co$_{78}$ with Ta under and over layers; it has in-plane magnetization which cannot be probed using polar MOKE, hence it's unknown if it exhibits HI-AOS.}

\end{figure*}

 A complete microscopic theory has yet to explain the intrinsic fundamental physics of HI-AOS. In this work the role of Gd in enabling HI-AOS was investigated experimentally and theoretically by studying \textit{a}-Gd$_{22-x}$Tb$_x$Co$_{78}$ thin films. Vapor deposited \textit{a}-RE-TM films have long been known to possess significant perpendicular magnetic anisotropy (PMA) leading them to have been considered and sometimes used as magnetic recording media for decades. PMA is intrinsic to the structure, a result of subtle structural ordering due to growth processes\cite{harris_structural_1992,hellman1992growth}. Tb has significant orbital angular momentum compared to Gd ($L$ = 3 compared to $L$ = 0) and greater spin-orbit coupling than Gd, leading to the large PMA and larger magnetic damping\cite{russek2002magnetostriction,radu2009laser}. By systematically replacing the Gd atoms with Tb atoms (such that the RE composition is kept constant), and by post-growth annealing, the anisotropy, magnetic damping and spin-orbit coupling (SOC) are modified, while the magnetization and the Curie and compensation temperatures are held constant. 
 
 HI-AOS was found in \textit{a}-Gd$_{22-x}$Tb$_x$Co$_{78}$ films up to x = 18, while \textit{a}-Tb$_{22}$Co$_{78}$ only demagnetized upon laser irradiation. The threshold or critical fluence of the pulse needed for switching increased linearly with increasing Tb concentration and the switching dynamics became slower, indicating increasing difficulties in the switching process with increasing Tb. Annealing \textit{a}-Gd$_{10}$Tb$_{12}$Co$_{78}$ (a high x hence high magnetic anisotropy film) at 300 \degree C for one hour reduced the anisotropy by an order of magnitude (without changing the magnetization), due to increased randomization of the local uniaxial anisotropy fields that lead to PMA. This then increases the damping and is here shown to yield slower remagnetization dynamics but no change in critical fluence. These slower dynamics with no change in critical fluence are reproduced in simulations with increased damping. Simulations of the atomistic spin dynamics using the \textsc{vampire} software package\cite{Evans2014,vampire-url} combined with a two-temperature model (2TM)\cite{ostler_ultrafast_2012} indicate that these results can be explained by an increased damping on the RE site when Gd is replaced by Tb. The results and simulations suggest that engineering the relative element-specific damping of the RE and TM sublattices can be used to find new materials that exhibit HI-AOS. Increased anisotropy with Tb content, and decreased anisotropy by annealing both led to slower switching dynamics, indicating that anisotropy is not a significant factor in HI-AOS. The high perpendicular magnetic anisotropy in Tb-rich films make them attractive candidates for magnetic devices with higher memory storage densities and retention times that exploit the fast read-write speeds of HI-AOS.

\section{Results}

\subsection{Single-shot HI-AOS in \textit{a}-GdTbCo alloys}

Amorphous, ferrimagnetic thin-films of Ta(3)/Pt(3)/\textit{a}-Gd$_{22-x}$Tb$_x$Co$_{78}$(10)/Pt(3) (thicknesses are in nm) heterostructures were sputter deposited onto substrates of Si(525$\mu$m)/SiO$_2$(50nm)/SiN$_x$(300 nm). The \textit{a}-Gd-Tb-Co films were co-deposited from separate Tb, Gd and Co targets with Pt or Ta over and underlayers grown \textit{in situ} (Methods section at the end). Energy dispersive spectroscopy images taken with a scanning transmission electron microscope found no evidence of inhomogeneities at the 10 nm scale as had been reported in previous work on \textit{a}-Gd-Fe-Co\cite{graves_nanoscale_2013} (See suppl. matls.). The magnetization was measured with a Quantum Design MPMS SQUID magnetometer as a function of field and temperature. Room temperature saturation M$_s$ ($\sim$100 emu/cm$^{3}$) and compensation temperature T$_M$ ($\sim$400 K) were independent of x, due to the fixed 22 at.\% RE content. The intrinsic perpendicular magnetic anisotropy constant K$_{ui}$ increased with increasing x, from 4 $\times$ 10$^{5}$ ergs/cm$^{3}$ for \textit{a}-Gd$_{22}$Co$_{78}$ to 2 $\times$ 10$^{6}$ ergs/cm$^{3}$ for \textit{a}-Tb$_{22}$Co$_{78}$.  At room temperature, the magnetization of all the films studied is RE-dominant. The Curie temperature of all films was found to be greater than 600 K.

The magnetization of the samples is initialized at room temperature with an external out of plane magnetic field of $\sim$ 0.7 T, fully saturating all samples. The field is then turned off, and the samples are irradiated with 100 fs full-width half maximum (FWHM) optical pulses from a regeneratively amplified Ti-Sapphire laser.  Magneto  Optical  Kerr  Effect  (MOKE) microscope images depict single shot switching of the magnetization of all films except \textit{a}-Tb$_{22}$Co$_{78}$ as  shown  in  Fig. 1a. Films with as much as 18 at.\% Tb (and hence as little as 4 at.\% Gd) have deterministic magnetization reversal upon irradiation with a single laser pulse. Amorphous Tb$_{22}$Co$_{78}$ only shows demagnetization, evidenced by the nucleation of random magnetic domains. Further growth and characterization details are available in supplemental materials.  

Fig. 1b shows that increasing the Tb content increases the incident critical fluence, starting from 4.4 mJ/cm$^2$ for \textit{a}-Gd$_{22}$Co$_{78}$ and linearly increasing to 6.2 mJ/cm$^2$ for \textit{a}-Gd$_4$Tb$_{18}$Co$_{78}$. The absorbed critical fluence calculated from ellipsometry measurements of the optical properties of the films (see suppl. matls.) increases linearly from 1.8 mJ/cm$^2$ to 2.5 mJ/cm$^2$ for \textit{a}-Gd$_{22}$Co$_{78}$ and \textit{a}-Gd$_{4}$Tb$_{18}$Co$_{78}$ respectively. 

The intrinsic anisotropy constant (K$_{ui}$) was measured at room temperature and is plotted as a function of Tb at.\% in Fig. 1c. The anisotropy constant increases systematically with increased Tb due to its large single ion anisotropy. The effect of growing on Ta or SiN, and of capping with Ta or Pt are also shown in Fig. 1c; these over and under layers, particularly Pt, increase the magnetic anisotropy of these thin films, nearly certainly due to interfacial anisotropy effects of these high SOC elements.  The HI-AOS was unaffected by these buffer layers, clear evidence that K$_{ui}$ alone is not the relevant driving parameter for HI-AOS.  We also note that several films of other RE/TM ratio were tested, with varying magnetization values M, including TM-dominant (low RE/TM ratio). These also exhibited HI-AOS, demonstrating that at least over a limited range of M, M is not a determining factor for the ability to show HI-AOS.

For $x=0$, i.e. \textit{a}-Gd-Co, a number of different types of films were prepared and measured, with different Gd/Co ratios and different K$_{ui}$, as shown in the legend and data of Fig 1c. With Pt over and underlayer, \textit{a}-Gd-Co has PMA and shows HI-AOS.  With Ta over and underlayer, \textit{a}-Gd-Co has in plane anisotropy and polar MOKE cannot measure HI-AOS. A Gd-Co-O sample (black diamond in Fig. 1c) was grown via oxygen reactive sputtering and exhibited both PMA and HI-AOS, as well as a distinct blue hue compared to the metallic gray of all other samples. The origin of PMA in \textit{a}-Gd-Co is still an open question\cite{bergeard2017correlation}, although this reactively-grown sample suggests that oxidation plays a significant role, and we have found that a-Gd-Co grown under very clean conditions and without Pt over or underlayers are magnetized in-plane, while slightly worse background pressure or deliberate O introduction yields PMA. Rutherford backscattering spectrometry (RBS) measurements of \textit{a}-Gd$_{22}$Co$_{78}$ of the main study do not reveal an oxygen peak, but its presence is likely hidden by the overlap of the Si signal from the substrate. Simulations with SIMNRA\cite{mayer1997simnra} indicate that if oxygen is present in \textit{a}-Gd$_{22}$Co$_{78}$ it is at most 5 at.\% O. Therefore it's likely that the PMA in \textit{a}-Gd-Co studied here is a combination of slight gettering of residual oxygen in the chamber, plus an interfacial contribution from heavy metal layers.

\subsection{Time dynamics of HI-AOS of \textit{a}-GdTbCo alloys}
The temporal dynamics of the magnetization of the \textit{a}-Gd$_{22-x}$Tb$_{x}$Co$_{78}$ films as they undergo HI-AOS was measured by time-resolved magneto optical Kerr effect (TR-MOKE) (see suppl. matls.). An incident fluence of 6.9 mJ/cm$^2$ was chosen for all TR-MOKE experiments as it slightly exceeds the incident critical fluence of \textit{a}-Gd$_4$Tb$_{18}$Co$_{78}$, the sample with the highest critical fluence that was able to be switched. Note that the dynamics are not affected by the fluence as shown in supplemental materials. Fig. 2a shows the ultrafast magnetization dynamics of all samples. The magnetization reversal process follows a two-step behavior. In the first step a rapid initial drop in the magnetization ocurs in which all films share a similar demagnetization process within the first picosecond post irradiation from the pump pulse. The second stage consists of remagnetization in the opposite direction as the system cools down, except for \textit{a}-Tb$_{22}$Co$_{78}$. \textit{a}-Gd$_{22}$Co$_{78}$ exhibits the fastest remagnetization time; with increasing Tb concentration, the remagnetization systematically slows. The remagnetization rate plateaus with 15\% and 18\% Tb samples exhibiting similar dynamics. Finally, \textit{a}-Tb$_{22}$Co$_{78}$ only demagnetizes upon irradiation and then recovers its magnetization along its initial direction upon cooling. It is possible that \textit{a}-Tb$_{22}$Co$_{78}$ exhibits a transient switching in the first few ps following irradiation, similar to the behavior reported by Alebrand et al.,\cite{alebrand_subpicosecond_2014} and modeled by Moreno et al.,\cite{moreno_conditions_2017}, suggesting that switching could ocur at a higher fluence.  However, utilizing higher fluences led to irreversible damage of the sample as the laser ablated or “burned” the sample surface. By 200 ps all samples had remagnetized to about 80\% of the saturation value as shown in supplemental materials. 

\begin{figure}[]
\centering
\includegraphics[width = 0.45\textwidth]{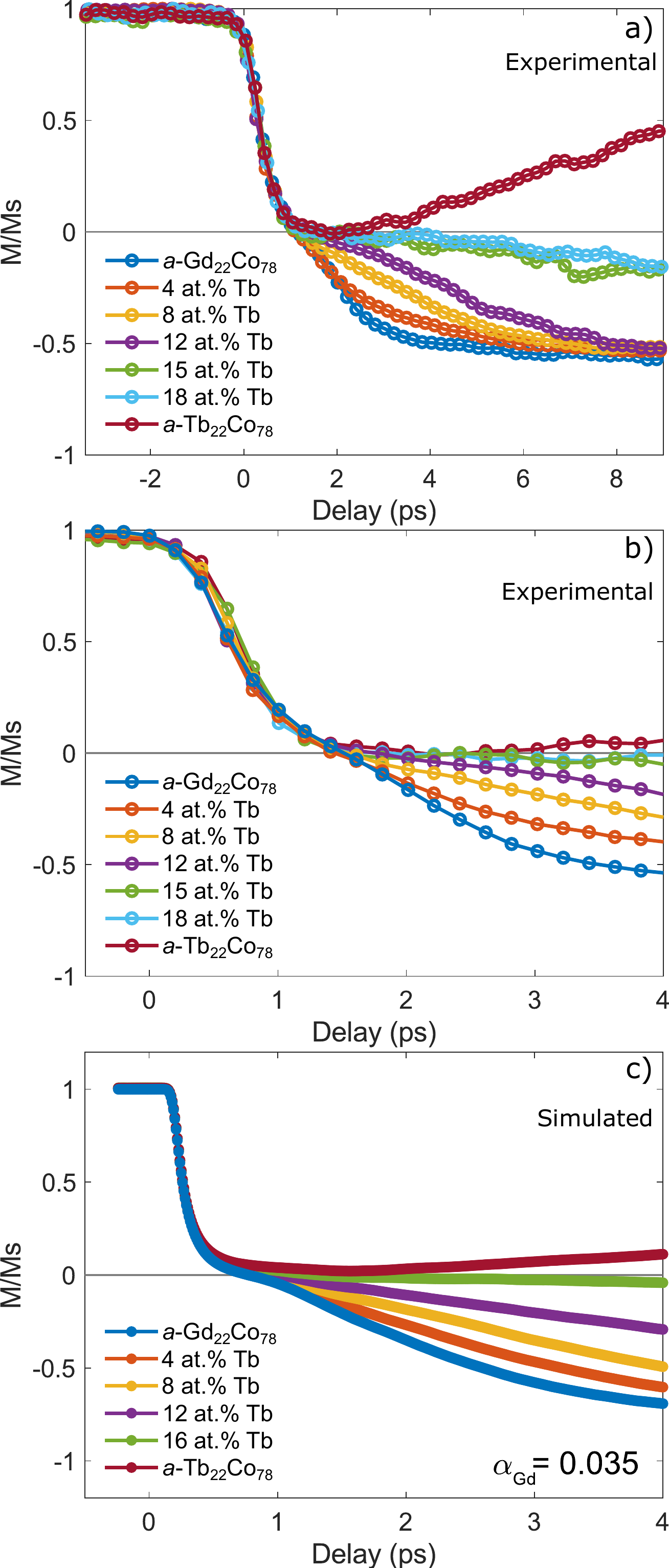}
    \caption{a) Time-resolved magnetization dynamics of a-Gd$_{22-x}$Tb$_x$Co$_{78}$ thin films following irradiation with laser pulses of 6.9 mJ/cm$^2$ fluence and 100 fs pulewidth. The initial rapid drop in magnetization is similar across all samples, but upon entering the remagnetization regime different rates are observed. b) Close up of the experimental data showing a bump in the magnetization (at $\sim$1 ps) following the initial demagnetization step. c) Simulation results of the magnetization dynamics of Gd$_{22-x}$Tb$_x$Co$_{78}$  after laser irradiation obtained with a two-temperature model neglecting spin-lattice coupling as described in the text. In this simulation, the damping coefficients for Tb and Co were 0.05\cite{moreno_conditions_2017} and the damping of Gd $\alpha_{Gd}$ was fixed at 0.035.}
\end{figure}

Fig. 2b is a close up of the experimental data of Fig. 2a, and shows, after the initial fast demagnetization step, behavior that deviates from exponential decay functions that appears to be more linear in character. The duration of this more linear behavior increases with increasing Tb before resuming exponential decay characteristics.

\subsection{Simulation of HI-AOS in \textit{a}-GdTbCo alloys}
Atomistic spin dynamics simulations using the \textsc{vampire} software package\cite{Evans2014,vampire-url} combined with a two-temperature model (2TM)\cite{ostler_ultrafast_2012} (details discussed in suppl. matls.) were performed to simulate the experimental magnetization dynamics. These simulations incorporate the Hamiltonian of the system, which includes the exchange and anisotropy energies of Gd, Tb and Co, into the Landau-Lifshitz-Gilbert equation to compute the dynamics. As shown in Fig. 2c the simulations are in excellent agreement with the experiments, reproducing the characteristic behavior of similar demagnetization dynamics followed by increasingly slow remagnetization times with increasing Tb content. The bump in the magnetization following the initial demagnetization step is clearly seen in simulation, and exhibits a more linear character with increasing Tb as seen experimentally. The approximately factor of two discrepancy in the time scales between experiment and simulation is due to both the small size of the simulated system, which does not allow for domain dynamics to be taken in consideration, and also due to heat dissipation effects.

The simulations were only able to reproduce the experimental data when the element specific damping of the Gd, Tb and Co sites were assigned separately, rather than when using a net damping of the system as is typically done when simulating such RE-TM systems \cite{iacocca2019spin,lu2018roles,moreno_conditions_2017} (see suppl. matls. for comparison). This is consistent with previous experimental and theoretical work on the role of damping in systems doped with RE\cite{radu2009laser,ellis2012classical}.
The experimental work of Radu et al. \cite{radu2009laser} on permalloy doped with RE showed increased damping for doping with Tb but no significant increase when doping with Gd. Ellis et al. \cite{ellis2012classical} showed that element-specific damping is required to reproduce the macroscopic damping in such systems.  
In this work we varied the Gd/Tb ratio, as in the experiments, and fixed the element-specific damping value at 0.05 of the Tb and Co sites as in ref. \cite{moreno_conditions_2017}.  The damping of Gd was taken to be lower than Tb and was varied between 0.005 and 0.05 in order to study its effect on the dynamics. In Fig. 2c, it is set at 0.035.

Fig. 3 shows the simulation results of critical fluence as a function of Tb concentration and varying Gd damping. The critical fluence for switching in the simulation is determined by the transition from non-deterministic to deterministic thermally induced switching. It shows that increasing the Tb content increases the critical fluence as observed experimentally, for all values of Gd damping. It also shows, that for a given concentration, increasing the damping on the Gd site increases the critical fluence.

\begin{figure}[]
\centering
\includegraphics[width = 0.45\textwidth]{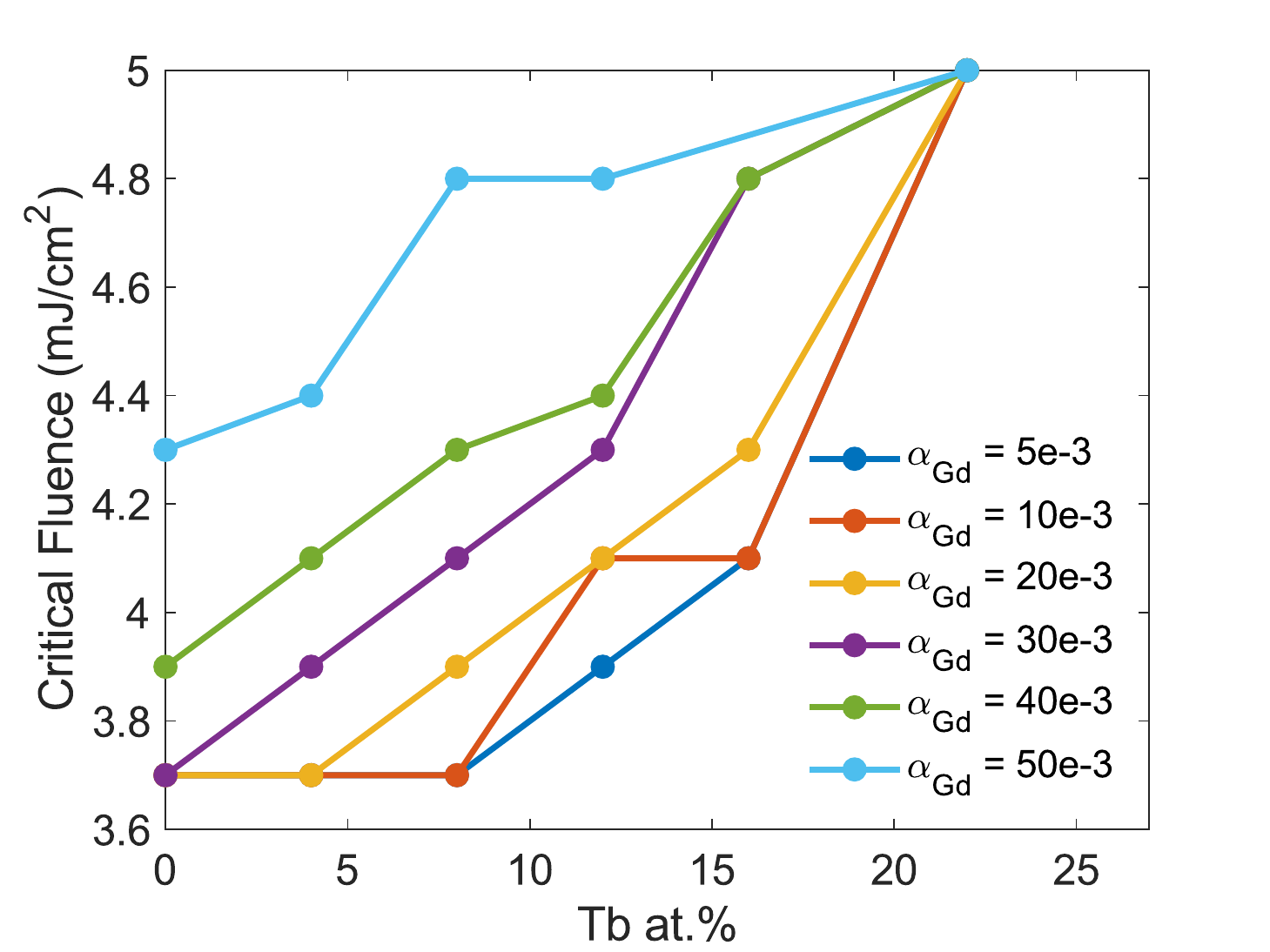}
  \caption{Simulated critical fluence of \textit{a}-Gd$_{22-x}$Tb$_{x}$Co$_{78}$ films as a function of Tb at.\% x for different damping values of Gd, $\alpha_{Gd}$. The Tb and Co damping parameters are constant (0.05). The critical fluence increases as $\alpha_{Gd}$ increases, indicating that lowering $\alpha_{Gd}$ reduces the threshold switching condition of \textit{a}-Gd-Tb-Co alloys below the laser ablation limit.}
\end{figure}

\subsection{Effect of annealing on switching dynamics}

Annealing was used to test the influence of anisotropy and damping on the dynamics; the results are shown in Fig. 4. Annealing \textit{a}-Gd$_{10}$Tb$_{12}$Co$_{78}$ at 300 $\degree$C for one hour results in a significant reduction in coercivity H$_c$  and anisotropy K$_{ui}$ (from 4.6 $\times$ 10$^{5}$ erg/cm$^3$ to 2.5 $\times$ 10$^5$ erg/cm$^3$) while maintaining the composition and M$_{\text{S}}$ constant as seen in Fig 4b. Further annealing at 350 $\degree$C eliminated the PMA. The fact that M$_{\text{S}}$ was unchanged by annealing strongly indicates that inhomogeneities such as phase segregation or crystallization have not ocurred. 
The annealed sample shows a significantly slower remagnetization time, as seen in Fig. 4b. Atomistic simulations of the magnetization dynamics of this sample as a function of Gd damping are shown in Fig. 4c, which shows that increased damping increases the remagnetization time. It can be seen that increasing the damping on the Gd sites explains the slower remagnetization time, suggesting that the experimentally annealed sample has increased damping, in addition to reduced anisotropy. Work from Malinowski\cite{malinowski_magnetization_2009} et al. showed that introducing local variations of the anisotropy in amorphous CoFeB leads to an increase in the damping parameter. Since the origin of anisotropy in RE-TM alloys is due to a combination of pair-ordering and Tb's single ion anisotropy\cite{harris_structural_1992,hellman1992growth}, annealing of \textit{a}-RE-TM alloys leads to a structural relaxation of pair-ordering that introduces local anisotropy variations. This in turn leads to higher damping and the slower remagnetization time observed. Fig. 2 showed that the samples with higher anisotropy (larger Tb at.\%) show slower switching dynamics in both experiments and simulation. But the annealing study in Fig. 4 shows that the film with lower anisotropy exhibits slower switching. These observations lead us to conclude that it is the damping of the system, and not the anisotropy, that is the significant contributor to the ability to exhibit HI-AOS, consistent with the observation made in connection with Fig. 1 that HI-AOS was not determined by the magnitude of K$_{ui}$. Simulations shown in Fig. 4c with varying damping support this conclusion. 

\begin{figure}[h!]
\centering
\includegraphics[width = 0.49\textwidth]{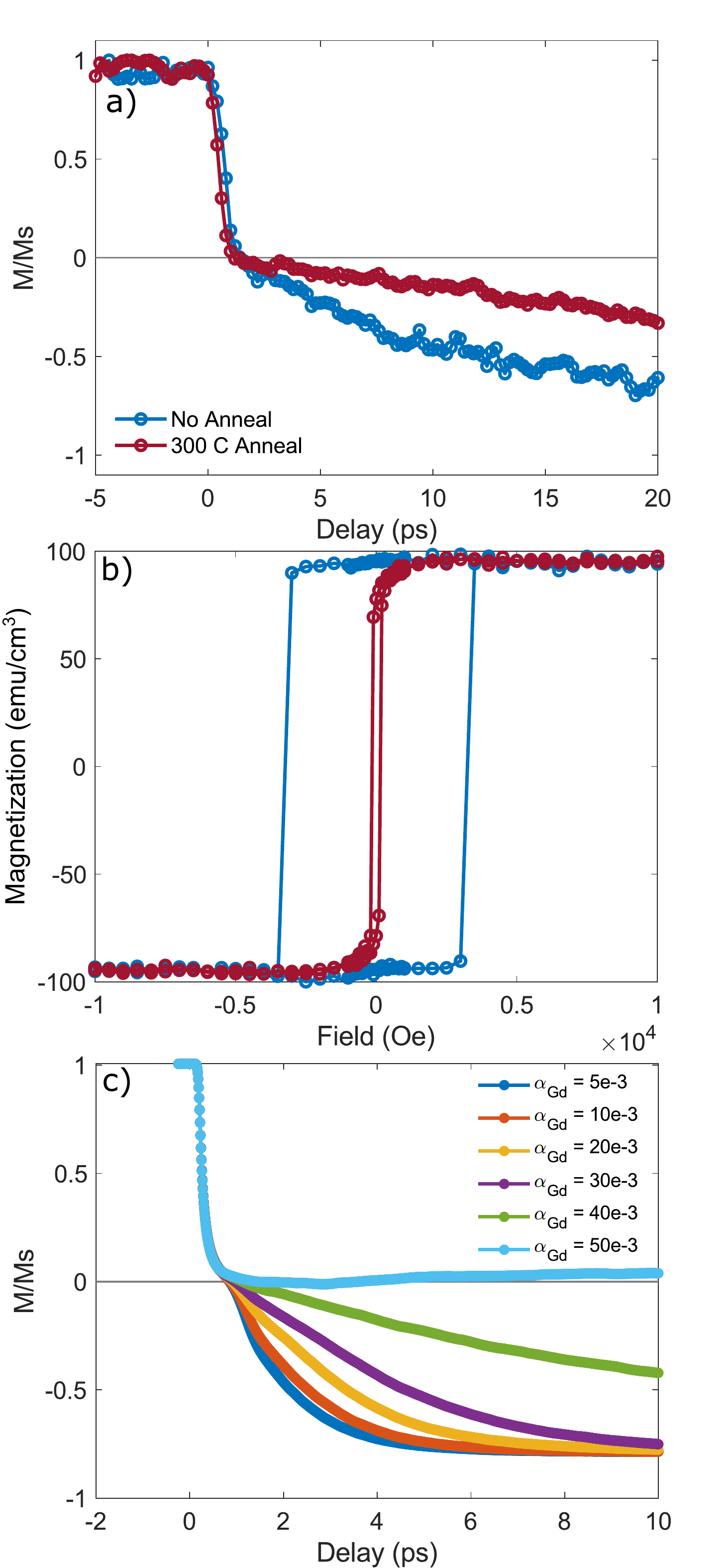}
  \caption{a) Magnetization dynamics of \textit{a}-Gd$_{10}$Tb$_{12}$Co$_{78}$ in the as-grown state (blue curve) and after annealing (red curve) at 300 $\degree$C for 1 hour. Annealing leads to a slower remagnetization time. b) Magnetization loops in the out-of-plane orientation show that annealing reduces the coercivity and anisotropy while M$_{\text{S}}$ is unchanged. c) Simulated time-resolved magnetization dynamics of \textit{a}-Gd$_{10}$Tb$_{12}$Co$_{78}$ as a function of increasing Gd damping. Increasing the damping leads to a slower remagnetization time, indicating that annealing leads to a higher damping value. }
\end{figure}

\section{Discussion}

The experimental results show increasing critical fluence and remagnetization times with increasing Tb content, while post-growth annealing slowed the remagnetization rates. The simulations strongly indicate that the relative element specific damping of the rare earth site compared to the cobalt site is the key factor that influences the critical fluence required switching and the speed of remagnetization. As verified in the model for the annealed sample, at a fixed composition the key parameter that leads to slower remagnetization is the increased elemental damping. In simulation the damping constant is a phenomenological parameter that combines a host of diverse effects.  Increasing the Tb composition leads to a greater spin-orbit interaction in the system, which is considered the intrinsic source of damping\cite{hickey_origin_2009} and is proportional to $\xi^2/W$, where $\xi$ is the spin orbital coupling energy and $W$ is the d-band width\cite{he_quadratic_2013}. Thus as the system becomes Tb-rich it experiences increased spin-orbit coupling which increases damping and thus leads to the dynamics observed in Fig. 2. For the dynamics observed in the annealed sample, the local spin-orbit coupling is very unlikely to have changed, but the anisotropy changed due to the structural relaxation and consequent randomization of local anisotropy axes induced by annealing, thereby increasing the macroscopic damping. The simulation does not directly acount for spin-orbit coupling, but it indirectly simulates the effects of stronger spin-orbit coupling via increases in the damping parameter. As mentioned before the deterministic switching is independent of the anisotropy and therefore the main contribution of spin-orbit coupling for all optical switching is the damping.  Therefore although the simulations reveal the critical role of damping in modifying the ultrafast magnetization dynamics, it is likely that the underlying physical mechanism is rooted in the spin-orbit interaction.

\section{Conclusion}
In conclusion, we have shown that \textit{a}-Gd$_{22-x}$Tb$_x$Co$_{78}$  thin films exhibit HI-AOS from x = 0 to x = 18, displaying a two-step reversal process with an identical fast demagnetization step and second slower remagnetization. The remagnetization time and the critical fluence increase as the Tb content increases, indicating that replacing Gd with Tb atoms hinders the switching process speed, but Tb helps increase PMA which is required for competitive memory storage technologies. Atomistic simulations explain the switching dynamics of our material system only after taking into acount the individual element specific damping of each RE and TM sites. The slower remagnetization dynamics and the increased critical fluence as the Tb atomic percentage increases are explained on the basis of increased damping on the RE site upon addition of Tb. Annealing of an \textit{a}-Gd$_{10}$Tb$_{12}$Co$_{78}$ film, which reduces K$_{ui}$ and Hc without changing M, led to a slowing of the remagnetization rate without changing the critical fluence; these results are captured in simulations which indicate that reduced damping is responsible for modifying the ultrafast magnetization dynamics. The experimental inability to observe HI-AOS in \textit{a}-TbCo is due to the high critical fluence that under the present conditions is inacessible without ablating the film. This suggests that better management of the laser thermal load may lead to HI-AOS beyond Gd-TM alloys. Our results indicate that the engineering of element specific damping of RE-TM systems will be crucial in endeavors to uncover material systems exhibiting HI-AOS with favorable properties (such as high anisotropy and large magnetization) for applications in ultrafast spintronic devices.

\section{Acknowledgments}
This work was primarily supported by the Director, Office of Science, Office of Basic Energy Sciences, Materials Sciences and Engineering Division, of the U.S. Department of Energy under Contract No. DE-AC02-05-CH11231 within the Nonequilibrium Magnetic Materials Program (KC2204). Ultrafast laser measurements were supported by the NSF Center for Energy Efficient Electronics Science. A.C. acknowledges support by the National Science Foundation under Grant No. DGE 1106400. This project has received funding from the European Union’s Horizon 2020 research and innovation programme under Grant Agreement No. 737093 (FEMTOTERABYTE). The atomistic simulations were undertaken on the \textsc{viking} cluster, which is a high performance compute facility provided by the University of York.  We are grateful for computational support from the University of York High Performance Computing service, \textsc{viking} and the Research Computing team.

A.C., A.P., A.E-G., F.H., and J.B. conceived and planned the experiments. A.C. and C.S. grew and characterized the samples. Laser measurements were performed by A.P. and A.E.-G. Simulations were performed by S.R. on the VAMPIRE software package developed by T.O., R.W.C. and R.F.L.E. TEM and EDS measurements were done by E.K. and M.S. A.C. and A.P. wrote the manuscript with input from all the authors.

\bibliography{hiaos.bib}

\end{document}